\documentclass{aastex}
\usepackage{emulateapj5}

\slugcomment{Accepted for Part 1 of the Astrophysical Journal}

\shorttitle{Jimenez-Garate et al.}
\shortauthors{X-ray Line Emission from Accretion Disk Atmospheres}

\begin{document}


\title{X-ray Line Emission from \\
 Evaporating and Condensing Accretion Disk Atmospheres}


\author{M. A. Jimenez-Garate\altaffilmark{1,3}, J. C. Raymond\altaffilmark{2},
D. A. Liedahl\altaffilmark{1}, C. J. Hailey\altaffilmark{3} }


\altaffiltext{1}{Lawrence Livermore National Laboratory, 
Department of Physics and Advanced Technologies, 7000 East Ave., L-41, Livermore, CA 94550;
liedahl1@llnl.gov}
\altaffiltext{2}{Center for Astrophysics, 60 Garden St., Cambridge, MA 02138;
raymond@cfa.harvard.edu}
\altaffiltext{3}{Columbia Astrophysics Laboratory, 538 West 120th St.,
New York, NY 10027; mario@alum.mit.edu, chuckh@astro.columbia.edu}


\begin{abstract}
We model the X-rays reprocessed by an accretion disk in a fiducial 
low-mass X-ray binary system with a neutron star primary. An atmosphere, or
the intermediate region between the optically thick disk and a 
Compton-temperature corona, is photoionized by the neutron star continuum. X-ray lines
from the recombination of electrons with ions dominate the atmosphere emission 
and should be observable with the \it Chandra \rm and \it XMM-Newton \rm high-resolution spectrometers.
The self-consistent disk geometry agrees well with optical observations of these
systems, with the atmosphere shielding the
companion from the neutron star.
At a critical depth range, the disk gas has one thermally
unstable and two stable solutions. A clear difference between the model spectra
exists between evaporating and condensing disk atmospheres.
This difference should be observable in high-inclination X-ray binaries,
or whenever the central continuum is blocked by absorbing material and
the extended disk emission is not. 
\end{abstract}



\section{Introduction}

The X-rays reprocessed on accretion disks to the optical, UV and X-ray bands
in both X-ray binaries and AGN could be one of our best probes of
accretion disk structure and the environment around compact objects.
It is convenient to study an illuminated disk around neutron star
low-mass X-ray binaries (LMXB) since the central continuum dominates the
energetics of the disk atmosphere and accretion disk corona (ADC),
and the system geometry is often determined.
Measurements of the optical magnitudes
and lightcurve amplitudes of LMXB yielded estimates of the angle subtended
by the disk of $\sim$ 12$\degr$ and of the average fraction of X-rays
reprocessed by the disk into the optical of $\lesssim 0.1$ \citep[]{dejong}. 
Models that seem to correctly describe the
accretion disk interior have been unable to produce disk geometries consistent
with the optical data \citep[]{dub99}. Analytical models assuming an
isothermal disk show better agreement with the data, yet produce illuminated disks that
are significantly thinner than expected \citep[]{vrtilek}.

A pure continuum incident on the gaseous disk will photoionize and heat
the gas. Radiative heating sustains a $\sim 10^7$ K ADC and a
smaller $\sim 10^5$-$10 ^6$ K atmosphere emitting X-ray lines on top of the
optically thick disk.
X-ray spectra of LMXBs during dips show
strong Comptonized emission plus an excess flux at 0.65
keV consistent with the same covering fraction \citep[]{church},
positing that ADC and disks may coexist.
The apparent size of the ADC and the disk may not
be equal \citep[]{church2000}, 
since the emission mechanisms have different radial dependence.
 \citet[]{nay99} 
modeled the reprocessed continuum and fluorescence emission from the
inner disk of AGN with an ADC,
and \citet[]{ko94} modeled X-ray spectra of disk annuli in an LMXB with ADC and normally
incident radiation.

	In this paper, we calculate the structure of an accretion disk illuminated by the
grazing radiation of a neutron star in an LMXB with a Compton-temperature
ADC, building on a model by \citet[]{ray93}.
This model, which we will detail in a future paper, includes
a more accurate calculation of the vertical and radial disk structure than
previous models, accounts for thermal instabilities, and incorporates
recent atomic model calculations that predict X-ray
spectra at a resolution superior to the \it Chandra \rm and \it XMM-Newton
\rm spectrometers. We obtain the first self-consistent 
flared disk geometry that matches the disk thickness expected
from optical data, and we predict that broad X-ray lines will be
observable and will provide, among other diagnostics,
evidence for photo-evaporation, which is central to the origin of
winds \citep[]{woods}.

\section{Model atmosphere}
We consider an LMXB with a $M_{*} = 1.4$ $M_{\sun}$ primary
radiating an Eddington luminosity bremsstrahlung continuum
with $T=8$ keV.  We use a set of fiducial system parameters for a bright LMXB,
so application to a particular source will require using the
observed X-ray continuum to improve accuracy.
The maximum radius of the centrally-illuminated disk is $10^{11}$ cm,
so the orbital period $\sim 1$ day. The minimum radius is $10^{8.5}$
cm, below which the omitted effect of radiation pressure will 
dominate.

We obtain the vertical disk atmosphere structure of each annulus by integrating the
hydrostatic balance and 1-D radiation transfer equations for a slab geometry:
\begin{equation}
\label{eq:hydro}
\frac{\partial P}{\partial z} = - \frac{G M_{*} \rho z }{r^{3}}
\end{equation}
\begin{equation}
\label{eq:rad1}
\frac{\partial F_{\nu}}{\partial z} = 
	- \frac{\kappa_{\nu} F_{\nu} }{\sin \theta}
\end{equation}
\begin{equation}
\label{eq:rad2}
\frac{\partial F_{\nu}^{d}}{\partial z} = 
	- \kappa_{\nu} F_{\nu}^{d} 
\end{equation}
while satisfying local thermal equilibrium:
\begin{equation}
\label{eq:thermal}
\Lambda (P, \rho,  F_{\nu}) = 0 
\end{equation}
where $P$ is the pressure, $\rho$ is the density, 
$F_{\nu}$ is the incident radiation field, $F_{\nu}^{d}$ is the
reprocessed radiation propagating down towards the disk midplane,
$z$ the vertical distance from the midplane,  $G$ the gravitational
constant, $\theta$ the grazing angle of the radiation on the disk,
$r$ the radius, and
$\kappa_{\nu}$ is the local absorption
coefficient. The difference between heating and cooling
$\Lambda$ includes Compton scattering, bremsstrahlung cooling,
photoionization heating, collisional line cooling, and recombination line
cooling from H, He, C, N, O, Ne, Mg, Si, S, Ar, Ca, and Fe ions.
We assume cosmic abundances \citep{allen}.
The code from \citet[]{ray93} 
computes the net heating and ionization equilibrium, Compton
scattering, and line scattering using escape probabilities.  A new
disk structure calculation simultaneously integrates Eqs.
\ref{eq:hydro}-\ref{eq:rad2} by the
Runge-Kutta method using an adaptive stepsize control routine with error
estimation, and Eq. \ref{eq:thermal} is solved by a globally convergent
Newton's method \citep[]{numrec}. At the ADC height $z_{cor}$, the equilibrium 
$T$ is close to the Compton temperature $T_{compton}$,
from which we begin to integrate downward 
until $T < T_{phot}(r)$. The optically thick part of the disk, with
temperature $T_{phot}$, is assumed to be vertically isothermal
\citep[]{vrtilek}. To get $T_{phot}$, we assume that the viscous
energy and illumination energy is all locally radiated as blackbody
radiation. That is, for $z_{phot} \ll r$ and $R_{*} \ll r$:
\begin{equation}
\label{eq:locdisk}
\sigma T_{phot}^{4} \simeq \frac{3 G M_{*} \dot{M} }{8 \pi r^{3} }
	 + \frac{(1 - \eta) L_{x} \sin \theta(r)}{4 \pi r^{2}}
\end{equation}
where $R_{*}$ is the neutron star radius, $\dot{M},L_{x}$ the
accretion rate and luminosity, $\sigma$ the Stephan-Boltzmann constant,
and $\eta = 0.9$ is the X-ray `albedo' derived from optical
observations \citep[]{dejong}, such that $1-\eta$ is the fraction of
X-rays absorbed at the photosphere.  The height at which the
integration ends is defined as the photosphere height $z_{phot}$.
Thus, the assumption is that for $z < z_{phot}$ viscous
dissipation dominates heating.

The boundary conditions at the ADC are set to 
$P(z_{cor})= \rho_{cor} k T_{compton} / \mu m_{p}$,
$F_{x}(z_{cor})=L_{x}/4 \pi r^2$, and
$F_{\nu}^{d}(z_{cor})=0$,
where $F_{x} \equiv \int F_{v}d\nu$, $k$ is the Boltzmann constant, and $\mu$ is the average particle mass
in units of the proton mass $m_{p}$. The boundary
conditions at $z_{phot}$ for $F_{\nu}$ and $F_{\nu}^{d}$ are set free, 
and we use the shooting method \citep[]{numrec} 
with shooting parameter $\rho_{cor}$ adjusted until we satisfy
$P(z_{phot})= \rho_{phot} k T_{phot} / \mu m_{p}$
at the photosphere.
Note $\rho_{phot}$ is the viscosity ($\alpha$) dependent density 
calculated for an X-ray illuminated \citet[]{ss73} disk.

We calculate $\theta(r)$ iteratively.
To get $T_{phot}$ self-consistently from Equation \ref{eq:locdisk}, we need 
$\theta(r)= \arctan (dz_{atm}/dr) - \arctan (z_{atm}/r) + \arctan (R_{*}/r)$.
We neglect the term $\propto R_{*}/r$, which is valid for
$r \gtrsim 10^{8.5}$ cm. 
After an initial guess for $z_{atm}(r)$, we define it as the height
where $F_{x}$ is attenuated by $1/e$.
We stop after $\theta(r)$ and $T_{phot}$ converge to $\lesssim 10$ \%.
Note $F_{x} \rightarrow 0$ for $z \lesssim z_{phot}$.
Since $z_{atm}$ is not physically determined, but it is bound by
$z_{atm}>z_{phot}$, we ran a second model calculating $\theta(r)$ from
$z_{phot}$, which we will use to estimate systematic errors due to the
1-D radiation transfer calculation.

\section{Thermal instability of the photoionized gas}

The disk atmosphere exhibits a thermally unstable region,
which has direct consequences for the disk structure and X-ray spectrum.
The origin of the thermal instability is well understood,
and its behaviour depends on metal abundances and
the local radiation spectrum \citep[]{hess}.

\begin{figure}
\epsscale{.6}
\plotone{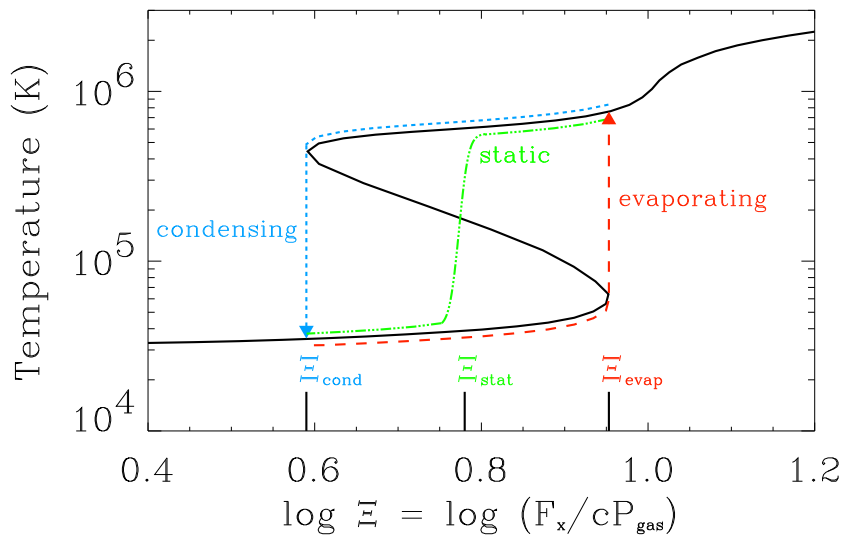}
\caption{Ionization parameter $\Xi$ vs. temperature
assuming an 8 keV bremsstrahlung continuum. The s-curve
corresponds to the locus of solutions in thermal balance
and ionization equilibrium. We model the extreme cases, the evaporating and
condensing disks, which correspond to the lower and upper stable branches
of the curve, respectively. The inclusion of conduction in the
energy equation modifies the thermal equilibrium locus, creating
solutions with one small transition region.  Schematics of the static
transition region at $\Xi_{stat}$ and two dynamic transition regions
at $\Xi_{cond}$ and $\Xi_{evap}$ are shown.  For $\log \Xi > 1.2$, $T$
increases up to $T_{compton} \sim 10^7$ K (not shown). \label{fig1}}
\end{figure}

Within a range of pressure ionization parameters ($\Xi$), 
thermal equilibrium is achieved by three distinct temperatures, only
two of which are stable to perturbations in $T$ (Fig. \ref{fig1}).
$\Xi \equiv 4 \pi P_{rad}/P_{gas}$, where $P_{rad} = F_{x} / 4 \pi c$ is
the radiation pressure and $P_{gas}$ the gas pressure \citep[]{kmt}. 
The thermally stable branches have positive slope \citep[]{field1965}.
Previous spectral studies of illuminated accretion disks in LMXB
had not explicitly selected the stable solutions \citep[]{ko94}, a choice
which affects X-ray production.

The instability implies a large temperature gradient 
as the gas is forced to move between stable branches, requiring
the formation of a transition region whose size may be determined by
electron heat conduction, convection, or turbulence,
depending on which dominates the heat transfer.
For simplicity, we neglect emission from the transition region.
Upon calculation of the Field length $\lambda_F$, 
the lengthscale below which conduction dominates
thermal equilibrium \cite[]{bemc}, we estimate that conduction 
forms a transition layer $\sim 10^{-2}$ times thinner than the
size of the X-ray emitting zones. Nevertheless, X-ray line emission
from the neglected conduction region may not be negligible in
all cases \citep[]{li}.

Conduction determines the dynamic behaviour of the gas. The dynamics
will allow us to attach a physical interpretation to the chosen solutions
from Fig. \ref{fig1}.  Roughly, a static conduction solution
can only be found if the transition region splits the instability
$\Xi$-range in half (Fig. \ref{fig1}). This solution \citep{zel} takes
the low-$T$ stable branch at $\Xi < \Xi_{stat}$, and the high-$T$
stable branch at $\Xi > \Xi_{stat}$, separated by the transition layer
at $\Xi_{stat}$. A transition layer located away from $\Xi_{stat}$
will dynamically approach $\Xi_{stat}$ by a conduction driven mass
flow.  A transition layer at $\Xi_{evap}>\Xi_{stat}$ is
moving towards lower $\Xi$, and the gas is evaporating from the low-$T$
branch to the high-$T$ branch.  A transition layer at
$\Xi_{cond}<\Xi_{stat}$ is moving towards higher $\Xi$, with the gas
condensing to the low-$T$ branch \citep[]{zel,li}.

We compute the disk structure for both condensing
and evaporating solutions. We assume a steady state,
condensing or evaporating mass flow through the transition layer at
$\Xi_{cond}$ or $\Xi_{evap}$, respectively. 
The static conduction solution is an
intermediate case of the latter extreme cases. 
We take a single-valued $T(\Xi)$, since a two-phase solution would be
buoyantly unstable, making the denser (colder) gas sink. 
The evaporating disk corresponds to the low-$T$ branch, while
the condensing disk corresponds to the high-$T$ branch (Fig. \ref{fig1}).
This introduces spectral differences (section \ref{sec:spec}). 

We do not know from first principles whether
the disk atmosphere is evaporating, condensing, or static.
We estimate the speed of the conduction mass flow to be $v_{cond} = 2
\kappa T / 3 P_{gas} \lambda_F$, by using the characteristic conduction
time at the Field length, where $\kappa$ is the \citet{spitzer}
conductivity \citep{mckee}.
We obtain a conduction mass flow speed $v_{cond}$ which is 1-2$\times
10^{-2}$ times the local sound speed. Thus, the phase dynamics will
depend on the subsonic ($v \gtrsim v_{cond}$) flow patterns in the
disk atmosphere, and these flows will also determine the evaporation
or condensation rates.

The only physical mechanism known to
transport the necessary angular momentum for disk accretion
involves a magneto-rotational instability (MRI) which drives turbulent
flow in the disk \citep[]{bh98}. Heat transfer due to this turbulent
flow could quench the thermal instability and affect the disk
structure. The MRI also favors our assumption of vertical
isothermality in the optically thick disk.  

\section{Spectral modeling}

With the disk structure $\rho(r,z)$, $T(r,z)$ 
and ion abundances $f_{Z,i}(r,z)$,
we model the X-ray line emission from
the disk atmosphere by using HULLAC (Hebrew University/Lawrence
Livermore Atomic Code, \citet[]{hullac}).
The code calculates the atomic structure and transition rates of
radiative recombination (RR) and radiative recombination continuum (RRC)
for the H-like and He-like ions of C, N, O, Ne, Mg, Si, S, Ar, Ca, and 
\ion{Fe}{17}-\ion{Fe}{26}. The emissivities are calculated
as described in \citet[]{sako}. 

The spectrum for each annulus is added to obtain the disk
spectrum.  Each annulus consists of a grid of zones in the vertical
$\hat{z}$ direction, and $T$, $\rho$ and $f_{Z,i+1}$ for each zone
are used to calculate the RR and RRC emissivities. The radiation 
is propagated outwards at inclination angle $i$, including
the continuum opacity of all zones above. 
Compton scattering is not included in the calculated spectrum 
because it is negligible in the vertical direction in the region 
where the line emission is formed. 
The spectrum is Doppler broadened by the projected
local Keplerian velocity, assuming azimuthal symmetry.

\section{Disk structure results}

The self-consistent disk is thicker than would be expected from
the local pressure scale height $H_{P}$ alone. 
We find  $H_{P} < z_{phot} < z_{atm}$. 
We fit the modeled photosphere and atmosphere height, $z_{phot}$ and
$z_{atm}$, with $Cr^n$ ($r$ in cm), with fit parameters $C$ and $n$. We find 
$C_{phot} = 2.4 \pm 0.4-$$1.9 \times 10^{-3}$ cm$^{1-n_{phot}}$, $n_{phot} = 1.14 \pm 0.01$+$0.06$,
$C_{atm} = 1.0 \pm 0.1$+$0.2 \times 10^{-3}$ cm$^{1-n_{atm}}$, $n_{atm} = 1.21 \pm 0.01$.
Thus, $z_{phot} \sim 3$-$4$ $H_{P}$ and $z_{atm} \sim 7$-$8$+$2$ $H_{P}$.
Estimated systematics are shown, if significant.
\citet[]{vrtilek} estimated $n_{atm} = 9/7=1.29$, but in spite of
the steeper radial dependence, the Vrtilek disk is thinner, and
it equals $H_{P}$ for $r > 10^{10}$ cm.
For comparison to the optically derived disk thickness, $z_{atm}$
is the \it de facto \rm disk boundary, since a fraction $1/e$ of the central
X-rays have been absorbed there. Thus, previous theoretical studies severely
underestimated the disk thickness.

The atmosphere's ($z > z_{phot}$)
maximum photoelectric opacity is always $\tau \ll 1$,
although $\tau / \sin \theta(r) \gtrsim 1$.
Illumination heating dominates at $r \gtrsim$ $10^{10}$ cm
(Eq. \ref{eq:locdisk}), where only
$F_{x}(z_{phot})/F_{x}(z_{cor}) \sim 0.12$ 
of the incident photons reach $z_{phot}$ directly, while
$F_{x}^{d}(z_{phot})/ ( F_{x}(z_{cor}) \sin \theta) \sim 0.35$ reach $z_{phot}$
after reprocessing in the atmosphere. Thus, the atmospheric albedo is
$\sim 0.5$.  To reproduce the disk (atmosphere and photosphere) albedo
observed by \citet[]{dejong} ($\eta \sim 0.9$), the photosphere albedo
can be $\lesssim 0.8$, which is a range closer to physical
expectations. There is no significant change in the atmosphere
structure by varying $\alpha$ from 1 to 0.1. 

\section{Spectral results}
\label{sec:spec}
The LMXRB photon flux (photon cm$^{-2}$ s$^{-1}$ keV$^{-1}$) is modeled by:
\begin{equation}
\label{eq:spec}
F_{\nu}^{tot} = e^{- \sigma_{\nu} N_{H}^{*}} F_{\nu}^{*} + 
e^{- \sigma_{\nu} N_{H}^{disk}} F_{\nu}^{disk}
\end{equation}
where $F_{\nu}^{*}$ is the neutron star continuum, 
$F_{\nu}^{disk}$ is the RR line and RRC modeled flux, 
$N_{H}^{*}$ and $N_{H}^{disk}$ are the neutral hydrogen
absorption column densities and $\sigma_{\nu}$ are the 
\citet[]{abun} absorption cross sections. The system is assumed to be
10 kpc away.

If we let $N_{H}^{*} = N_{H}^{disk}$ 
in Eq. \ref{eq:spec} with $i=0$, the lines are 
swamped by the continuum. Thus, low inclination 
neutron star LMXBs are unlikely to have detectable
X-ray lines from the disk. Instead, we consider 
$N_{H}^{*} = 5 \times 10^{22}$ cm$^{-2}$ and $N_{H}^{disk} = 10^{21}$
cm$^{-2}$ with $i=75^{\circ}$, where an
obscuring medium absorbs half of the continuum flux and
is compact enough to leave the disk almost unobscured.

\begin{figure}
\epsscale{.9}
\plotone{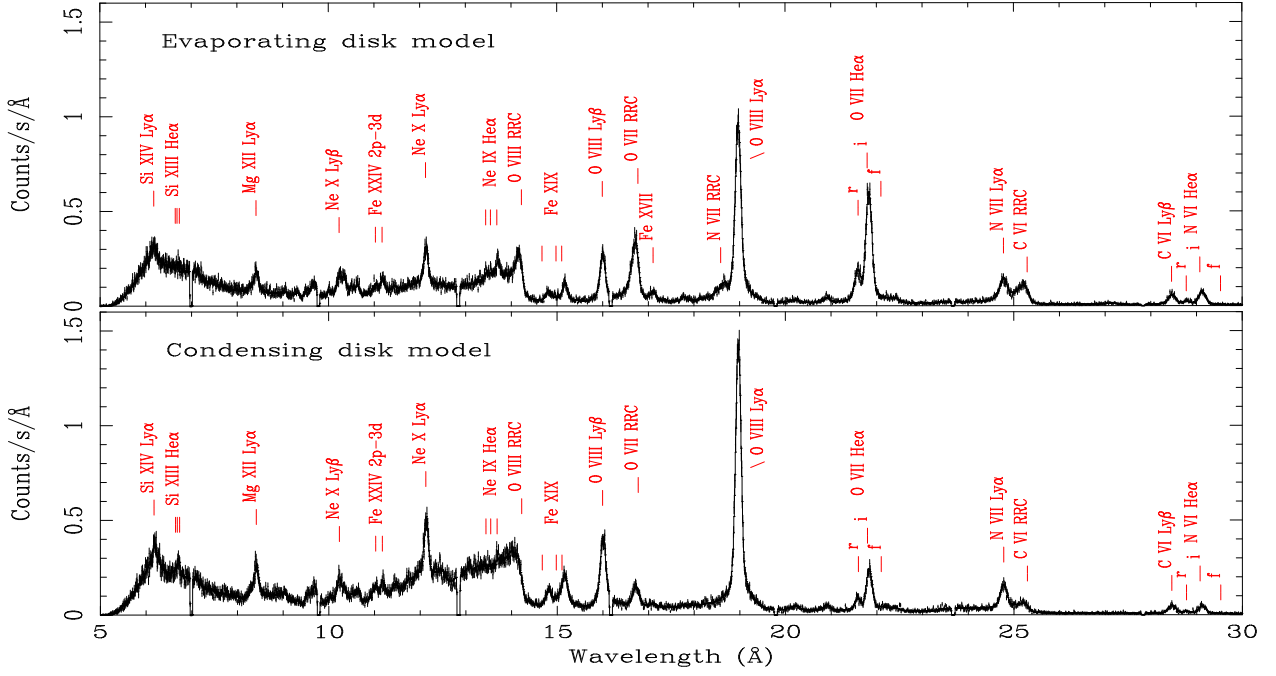}
\caption{Spectra for evaporating and condensing accretion
disks. Simulated 50 ks observation with \it XMM-Newton \rm RGS 1. 
The continuum emission from the inner ($r < 10^{8.5}$ cm)
disk is not included, and the obscuration of the ionizing continuum
from the compact object is assumed.
\label{fig2}}
\end{figure}

\begin{figure}
\epsscale{.73}
\plotone{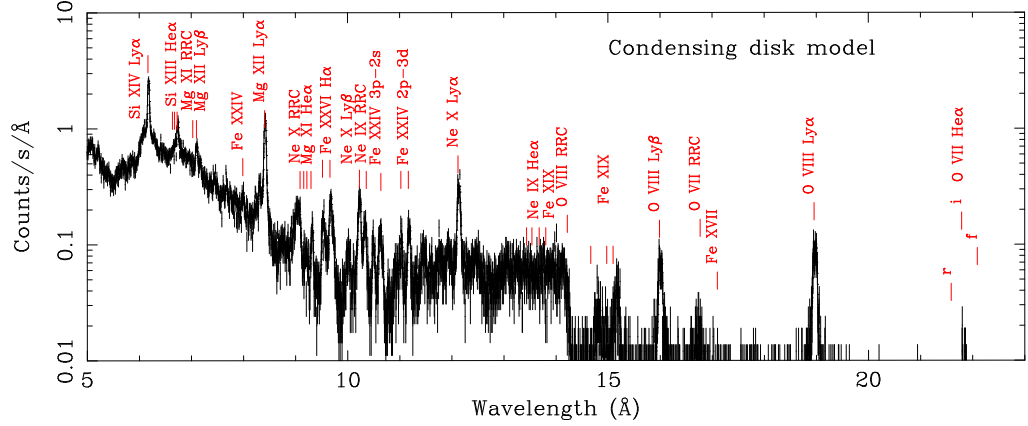}
\caption{Condensing disk spectrum (as in Fig.
\ref{fig2}). Simulated 50 ks observation with \it Chandra \rm MEG 1.
\label{fig3}}
\end{figure}

With a partially obscured central continuum, disk evaporation has an
observable spectral signature. We simulate 50 ks \it XMM-Newton \rm
RGS 1 and \it Chandra \rm MEG +1 observations (Fig. \ref{fig2},\ref{fig3}).
Some bright lines are shown on Table \ref{tbl-1}.
The evaporating and condensing disks
have contrasting \ion{O}{7}/\ion{O}{8} line ratios. The evaporating
disk contains gas at $T \sim 7$-$10 \times 10^4$ K, unlike the condensing
disk. The H-like ion line intensities are higher for the condensing disk
since it has more gas at $T \sim 10^6$ K.
The spectral differences stem from the distinct differential emission measure
distributions $d(EM)/d\log \Xi$ and from the \ion{O}{7} recombination
rate $\alpha_{RR} \propto T^{-\gamma}$, where
$\gamma = 0.7$-$0.8$, $EM=\int n_{i} n_{e} dV$ is the emission measure, and
$n_{i}, n_{e}$ are the ion and electron densities.

The RRC and He-like line triplet can be used as
temperature and density diagnostics, respectively.
The RRC width $\propto T$ \citep[]{lied}.
The forbidden line $f$ of \ion{O}{7} at $22.097$  
\AA \ is absent due to
collisional depopulation at high density, since 
$\rho \gtrsim 10^{13}$ cm$^{-3}$.
The \ion{O}{7} intercombination $i$ to resonance $r$ line  
ratio is $g=(f+i)/r \simeq i/r \gtrsim 4$, 
indicating a purely photoionized plasma \citep[]{porquet}.
The \ion{O}{7} RRC broadening is Doppler-dominated,
resembling the RR lines, while the \ion{O}{8} 
RRC has a component with $FWHM \sim 2$
\AA \ that is produced at $T \sim 10^6$ K.

\section{Conclusions}

Using a model for a disk illuminated by a neutron star, we 
have produced a global spectrum of X-ray line emission from a
self-consistent accretion disk atmosphere. The flaring of the atmosphere
matches optical observations \citep[]{dejong}. 
For fiducial parameters of a LMXB with a partially obscured central
source with Eddington luminosity, observable spectral signatures would
identify an evaporating disk atmosphere.

The \ion{O}{7}/\ion{O}{8} line ratios 
are a good tracer of evaporation from the disk. The line-rich
spectrum has line profiles resolvable by the \it XMM \rm and
\it Chandra \rm spectrometers.
The lines have broad wings and a double-peaked core
with a velocity from the outermost radius. Only a single 
line may be resolved since the emission is well distributed
throughout $10^{8.5}$ cm $< r < 10^{11}$ cm.
He-like triplet diagnostics show signatures of photoionization
and high density ($n_{e} > 10^{13}$ cm$^{-3}$), and H-like RRC are very broad, 
with a superposed narrow component in evaporating disks.

The distinct spectral signature of evaporation is due to the presence
of a thermal instability in the photoionized gas of the disk.
This instability cannot be quenched by spectral variations, abundance
variations, or optical depth effects \citep[]{hess}. Conduction heating 
should not affect this result, though turbulent heat transfer or
coronal heating might quench the instability if present. In the outer
disk, the incident radiative power is $\gtrsim 10$ times the viscous
power, so coronal-like heating should be small.

The X-ray lines from the disk are weak and 
difficult to detect for a low-inclination LMXB. Only if
the inclination is high and a cloudy environment exists do we
expect them to be detectable. At press time, an \it XMM-Newton \rm
observation of LMXB dipper EXO 0748-67 has detected broad lines
from photoionized gas with $n_{e} > 7 \times 10^{12}$ cm$^{-3}$
\citep[]{cottam}, which we interpret as the disk atmosphere emission,
since it confirms the key predictions of our model.

\acknowledgments{We are grateful to S. Kahn and I. Joseph for 
fruitful discussions.  This work was performed under the auspices of
the U.S. Department of Energy by the University of
California Lawrence Livermore National Laboratory under contract
No. W-7405-Eng-48.}





\begin{deluxetable}{crr}
\tabletypesize{\footnotesize}
\tablecaption{Line fluxes for two disk models. \label{tbl-1}}
\tablewidth{0pt}
\tablehead{
\colhead{Line(s)} & \colhead{Evaporating\tablenotemark{a}}   & \colhead{Condensing\tablenotemark{a}} 
}
\startdata
\ion{O}{7} He$\alpha$ triplet & 2.1 & 0.99 \\
\ion{O}{8} Ly$\alpha$ & 3.4 & 4.7 \\
\ion{O}{7} RRC & 1.3  & 0.61 \\
\ion{O}{8} Ly$\beta$ & 0.94 & 1.5 \\
\ion{Fe}{19} L (0.815 keV) & 0.45 & 0.91 \\
\ion{Ne}{10} Ly$\alpha$ & 1.4 & 2.2 \\
\ion{Fe}{26} H$\alpha$ (1.28 keV) & 0.74 & 0.86 \\
\ion{Mg}{12} Ly$\alpha$ & 1.2  & 2.0 \\
\ion{Si}{13} He$\alpha$ triplet & 1.2  & 2.4 \\
\ion{Si}{14} Ly$\alpha$ & 4.0 & 6.0 \\
\ion{Fe}{25} Ly$\alpha$ & 15 & 16 \\
\ion{Fe}{26} Ly$\alpha$ & 12 & 12 \\
\enddata
\tablenotetext{a}{Line fluxes in units of $10^{-12}$ erg cm$^{-2}$ s$^{-1}$,
typically over an interval $E/\Delta E \sim 100$. Estimated 
systematic normalization error limit of $\sim 50$ \%, due to the 1-D
calculation.}
\end{deluxetable}


\begin{thebibliography}{}
\bibitem[Allen(1973)]{allen} Allen, C.\ W.\ 1973, Astrophysical
Quantities (3rd ed.; London: Athlone Press)  
\bibitem[Balbus \& Hawley(1998)]{bh98} Balbus, S. A., \& Hawley, J. F. 1998,
	Rev. Mod. Phys., 70, 1
\bibitem[Begelman \& McKee(1990)]{bemc} Begelman, M. C., \& McKee, C. 1990,
	\apj, 358, 375
\bibitem[Church (2000)]{church2000} Church, M. J. 2000, astro-ph/0012411
\bibitem[Church et al.(1998)]{church} Church, M. J., Balucinska-Church, M.,
	Dotani, T., \& Asai, K. 1998, \apj, 504, 516
\bibitem[Cottam et al.(2001)]{cottam} Cottam, J., Kahn, S.\ 
M., Brinkman, A.\ C., den Herder, J.\ W., \& Erd, C.\ 2001, \aap, 365, L277 
\bibitem[de Jong, van Paradijs, \& Augusteijn(1996)]{dejong} de Jong, J. A., 
	van Paradijs, J., \& Augusteijn, T. 1996, \aap, 314, 484
\bibitem[Dubus et al.(1999)]{dub99} Dubus, G., Lasota, J. P., Hameury, J. M.,
	\& Charles, P. 1999, \mnras, 303, 139
\bibitem[Field(1965)]{field1965} Field, G.\ B.\ 1965, \apj, 142, 
531 
\bibitem[Hess, Kahn, \& Paerels(1997)]{hess} Hess, C. J., Kahn, S. M., \& Paerels, F. B. S.
	1997, \apj, 478, 94
\bibitem[Klapisch et al.(1977)]{hullac} Klapisch, M., Schwab, J. L.,
	Fraenkel, J. S., \& Oreg, J. 1977, Opt. Soc. Am., 61, 148 
\bibitem[Ko \& Kallman(1994)]{ko94} Ko, Y., \& Kallman, T. R. 1994,
 \apj, 431, 273
\bibitem[Krolik, McKee \& Tarter(1981)]{kmt} Krolik, J. H., McKee, C. F., \&
 Tarter, C. B. 1981, \apj, 249, 422
\bibitem[Li et al.(2001)]{li}Li, Y., Gu, M. F., \& Kahn, S. M.,
submitted to \apj
\bibitem[Morrison \& McCammon(1983)]{abun} Morrison, R., \& McCammon, D. 
	1983, \apj, 270, 119
\bibitem[Liedahl \& Paerels(1996)]{lied}Liedahl D. A., \& Paerels F. 1996,
		\apj, 468, 33
\bibitem[for a detailed discussion, see McKee \& Begelman(1990)]{mckee} McKee, C.\ F.\ \& 
Begelman, M.\ C.\ 1990, \apj, 358, 392 
\bibitem[Nayakshin, Kazanas \& Kallman(2000)]{nay99} Nayakshin, S., Kazanas, D., \&
Kallman, T. R. 2000, \apj, 537, 833
\bibitem[Porquet \& Dubau(2000)]{porquet} Porquet, D., \& Dubau, J. 2000,
	\aap, 143, 495
\bibitem[Press(1994)]{numrec} Press, W. H. 1994,  Numerical Recipes
in FORTRAN : The Art of Scientific Computing (Cambridge: Cambridge
University Press)
\bibitem[Raymond(1993)]{ray93} Raymond, J. C. 1993, \apj, 412, 267
\bibitem[Sako et al.(1999)]{sako} Sako M., Liedahl, D. A., Kahn, S.
M., \& Paerels, F. 1999, \apj, 525, 921
\bibitem[Shakura \& Sunyaev(1973)]{ss73} Shakura, N. I., \& Sunyaev, R. A.
	1973, \aap, 24, 337
\bibitem[Spitzer(1962)]{spitzer} Spitzer, L.\ 1962, Physics of 
Fully Ionized Gases, (New York: Interscience)  
\bibitem[Vrtilek et al.(1990)]{vrtilek} Vrtilek, S. D., Raymond, J. C., 
	Garcia, M. R., Verbunt, F., Hasinger, G., \& Kurster, M. 1990,
	\aap, 235, 162
\bibitem[Woods et al.(1996)]{woods} Woods, D. T., Klein, R. I., Castor, J. I., McKee,
 	C. F., \& Bell, J. B. 1996, \apj, 461, 767
\bibitem[Zeldovich \& Pikelner (1969)]{zel}Zeldovich, Y. B., \& Pikelner, 
	S. B. 1969, Soviet Physics JETP, 29, 170

\end{thebibliography}
\end{document}